# Towards the Objective Speech Assessment of Smoking Status based on Voice Features: A Review of the Literature

*Zhizhong Ma[a], Chris Bullen[b]\*, Joanna Ting Wai Chu[b], Ruili Wang[a], Yingchun Wang[a], Satwinder Singh[a]*

[a] *School of Natural and Computational Sciences, Massey University, Auckland, NZ*
[b] *National Institute for Health Innovation, University of Auckland, Auckland, NZ*

## Abstract

*Background and Objective*: In smoking cessation clinical research and practice, objective validation of self-reported smoking status is crucial for ensuring the reliability of the primary outcome, that is, smoking abstinence. Speech signals convey important information about a speaker, such as age, gender, body size, emotional state, and health state. We investigated (1) if smoking could measurably alter voice features, (2) if smoking cessation could lead to changes in voice, and therefore (3) if the voice-based smoking status assessment has the potential to be used as an objective smoking cessation validation method.

*Methods*: A systematic review of the scientific literature was conducted to compile studies on smoking status assessment based on voice features. We searched nine scientific databases for original studies involving the effects of smoking on voice features, the effects of smoking cessation on voice features.

*Results*: A total of 34 studies were identified for review. We found that fundamental frequency, jitter, shimmer, harmonics to noise ratio, and other voice features are affected by smoking and could be used to assess smoking status.

*Conclusion*: Speech assessment of smoking status based on voice features has potential as a smoking status validation method, as it is simple, reliable, and less time-consuming. Furthermore, this study provides recommendations for future research on the objective speech assessment of smoking status based on voice features.

---
\* This is to indicate the corresponding author.
  Email address: c.bullen@auckland.ac.nz

# 1. Introduction

Smoking remains one of the leading preventable causes of illness worldwide[1]. Conversely, stopping smoking (smoking cessation) dramatically reduces the risks of future disease and premature death. In smoking cessation research, the proportion of smokers who remain abstinent from smoking for a sustained period of time (typically measured at six months after the date of quitting) is the primary outcome measure used to evaluate the effectiveness of new interventions, policies, and practices for reducing smoking[2,3]. It is regarded as best practice in clinical trials and in the clinical treatment of smoking that self-reported abstinence should be confirmed with biochemical verification to validate smoking status. Several biological markers can be used to validate smoking status using biological samples. Different biological means of validating smoking status will be suitable in different studies, depending on the sample collection methods available, exclusion/inclusion of alternative nicotine delivery, the relevant timeframes for assessing smoking status, and availability of technical expertise. A consensus of experts in the field of smoking cessation suggests that carbon monoxide measures in expired breath and cotinine assay measured in urine or saliva are the most valid and feasible methods to validate smoking status[4], due to their relatively low cost and ease of use compared to other validation methods (Table 1).

Table 1. Comparison of biochemical smoking status validation methods.

| Validation Method | Type of sample | Maximum detection Window | Advantages | Disadvantages |
|---|---|---|---|---|
| Nicotine | Blood, saliva, urine | 8-12 hours | High specificity, samples can be sent for testing. | Short half-life, expensive, technical difficulty, not suitable for Nicotine Replacement Therapy (NRT) trials. |

| Carbon Monoxide | Expired air, blood | 12-24 hours | Relatively inexpensive, commercially available instruments, simple, portable, immediate feedback. | Marginal utility for light cigarette use, sensitive to environmental sources of CO (i.e. pollution, Second Hand Smoke, marijuana use), short detection window, in person testing, not suitable for non-combustible tobacco products. |
| --- | --- | --- | --- | --- |
| Cotinine (Nicotine metabolite) | Blood, saliva, urine | 80-100 hours | High specificity, longer half-life, inexpensive, can be assessed remotely (i.e. saliva test-strips). | Not suitable for NRT trials, biohazard risk constraints for collection and carriage of a specimen, half-life varies across groups |
| Minor tobacco alkaloids (anabasine, anatabine) | Urine | 50-80 hours | High specificity, differentiates NRT. | Expensive, technical difficulty. |

However, for some studies, the available biological methods for validating smoking status may not be feasible due to costs, remoteness, and sample collection. In such cases, self-report measures are widely used as alternative methods for estimate information concerning the smoking status[5,6]. However, self-report measures can be subject to bias and misreporting and may compromise the validity of the study findings and affect the trial outcome. Specifically, self-report measures in smoking cessation trials may be subject to social desirability bias, and participants may be biased to report non-smoker status because of a desire to 'help' the researcher be a 'good' participant and avoid stigma[7]. Research suggests that there is a high rate of misclassification for self-reported abstinence, which varies across populations[8,9]. Due to demand characteristics, misclassification may be highest among ethnic minority groups and lower socioeconomic groups, where the burden of harm from tobacco is highest[10,11].

As such, there is a need to develop new methods of validating smoking status that are simple, non-invasive, low-cost, able to be used across a widely dispersed population and do not require face-to-face contact. The development of such methods would revolutionise the way that smoking cessation studies are evaluated and population smoking trends monitored, improving the feasibility of smoking status validation, particularly in large studies.

Speech signals convey a speaker's important information such as age, gender, body size, emotional state, and health state[12–16]. The physiological effects of cigarette smoking have been well-documented[17–19] and include pharyngeal diseases and disorders resulting from prolonged effects of a variety of harmful chemicals in the cigarette smoke. Exposure to tobacco smoke can affect throat tissues, causing inflammation to vocal-folds and misfunction of the vocal cords[20–22], along with degrading lung function and thereby decreasing the airflow through the vocal cords[21,23,24]. The changes in the vocal tract can eventually lead to a dramatic variation in the speaker's speech signals. This raises the possibility that smoking status validation via speech analysis (as well as smoking behaviour detection) could be used to identify if a person is a smoker from a given speech signal by comparing his/her voice features with other smokers' and/or non-smokers' voice features. Advances in speech signal processing and machine learning could enable such analyses to be done in real time. There has been a set of research works on machine learning approaches applied to the broad voice analysis research area such as pathological voice detection[25,26], voice activity detection[27,28]. A number of studies have investigated voice analysis based on specific machine learning algorithms such as decision trees[29], support vector machine (SVM)[30,31], hidden Markov model (HMM)[32,33], Gaussian mixture model (GMM)[34,35], artificial neural networks (ANN)[36,37] and have reported high accuracy and performance[38–40].

We hypothesised that when a person changes their smoking behaviour, the change in their voice could be used to validate smoking status. The idea of identifying a smoker from a given speech signal by comparing his/her voice features with other smokers' and/or non-smokers' voice features has been explored in previous studies that concluded there is a correlation between a smoker's speech signals and his/her smoking status but were limited by a number of methodological features. Yet the question remains regarding: what kind of voice features will be affected by smoking? Voice features can be divided into two categories: linguistic features and acoustic features[41].

While linguistic features involve the analysis of language form, language meaning, and language in context, acoustic features are the acoustic components present in a speech that are capable of being experimentally observed, recorded, and reproduced; they include fundamental frequency ($F_0$), pitch, jitter, shimmer, harmonics to noise ratio (HNR). We were interested in the acoustic features that enable the analysis of speech signals because these features contain speakers' discriminative information that can be extracted for further classification.

## 2. Methods

In this research, we aimed to evaluate the effects of smoking and smoking cessation on acoustic voice parameters. We sought to answer the following four questions:

(1) Does smoking affect a speakers' voice quality?

(2) Which voice features are altered due to smoking and by how much?

(3) After cessation of smoking, do these features recover to a normal level, and over what period?

(4) Is it possible to detect if a person is an active smoker or an ex-smoker from these voice changes?

The following digital databases below were searched for relevant articles:

- ACM Digital Library [http://dl.acm.org]
- IEEE eXplore [http://ieeexplore.ieee.org]
- ScienceDirect [https://www.sciencedirect.com]
- SpringerLink [https://link.springer.com/]
- MDPI (Multidisciplinary Digital Publishing Institute) [https://www.mdpi.com/]
- arXiv [https://arxiv.org/]
- Taylor & Francis Online [https://www.tandfonline.com/]
- PubMed [https://pubmed.ncbi.nlm.nih.gov/]
- Google Scholar [https://scholar.google.com/]

The following search terms were used and linked: (smoking OR cigarette smoking OR tobacco smoking) AND (voice features OR voice parameters OR speech signal) AND/OR (voice analysis OR assessment) AND/OR (smoking cessation OR quit smoking). The scope of the study was restricted from any period to 2020.

Initially, a total of 17 papers were found. We then widened the search with additional keywords (smoker detection, voice analysis, acoustic analysis, etc.), leading to a total of 34 articles included for review. The following section summarises the key findings from the review in relation to the research questions.

## 3. Results

### 3.1 The Effects of Smoking on Voice Features

In the following sections, we list the voice features of a speech signal known to be affected by smoking.

#### 3.1.1 Fundamental frequency ($F_0$)

The fundamental frequency ($F_0$) is an important acoustic feature of speech signals. $F_0$ is the lowest, and usually, the strongest frequency produced by the complex vocal fold vibrations, measured in Hertz (Hz). It is generally considered to be the fundamental tones of sound and represents how high or low the frequency of a person's voice sounds. The $F_0$ is calculated by using the period $T$ of the speech signal:

$$F_0 = \frac{1}{T} \tag{1}$$

However, for the speech signal, the period $T$ is not constant since the input speech signal contains amplitude and frequency perturbations[42]. Several improved algorithms such as RAPT[43], SWIPE[44], YIN[45], and pYIN[46] have been proposed to estimate the $F_0$ based on acoustic features. Fundamental frequency values obtained in speech signals are typically less than 300 Hz for children and greater than 100 Hz for adults, 120 Hz for men and 210 Hz for women[47–49]. Studies have consistently found lower $F_0$ in smokers compared to age- and sex-matched non-smokers: in a study of 80 participants aged 25-49 years, half of whom were smokers[50], $F_0$ was measured for oral reading and spontaneous speech. The mean $F_0$ values were lower in the smoker group than the non-smoker group for males (105.65 Hz smokers vs 115.95 Hz non-smokers) and females (182.70 Hz smokers vs 186.45 Hz non-smokers). Gonzalez and Carpi evaluated the effect of cigarette smoking on voice features in young adults (n=134) who smoked less than 10 years[20]. $F_0$ was lower in smokers than in non-smokers, but the difference was only statistically significant in females (192.4 Hz smokers vs 206.4 Hz non-smokers, p<0.01). There was a dose-response effect with the number of cigarettes smoked. The findings suggest that the effect of smoking on $F_0$ is more significant for women than for men.

Lee et al. compared the voice features of non-smoking women that were exposed to second-hand smoke (i.e. passive smoking) to those that were not exposed[51]. There was no significant difference between passive smokers and non-smokers in $F_0$ (229 Hz passive smokers vs 234 Hz non-smokers), or for any of the other voice features.

Table 2 Comparison of smokers' vs non-smokers' mean fundamental frequency (Hz)

| Author(s) | Methods | Smokers | | Non-smokers | |
|---|---|---|---|---|---|
| *Horii and Sorenson. (1982)* | Oral reading | Male | Female | Male | Female |
| | Average | 105.65 | 182.70 | 115.95 | 186.45 |
| | 25-32 | 114.62 | 189.93 | 123.27 | 199.58 |
| | 33-41 | 106.71 | 197.67 | 118.49 | 178.32 |
| | 42-49 | 95.76 | 159.88 | 107.42 | 208.11 |
| *Gonzalez and Carpi (2004)* | Sustained vowels | Male | Female | Male | Female |
| | | 119.4 | 192.4 | 125.4 | 206.4 |
| *Lee et al. (1999)* | Sustained vowels | Female | | Female | |
| | | 229 | | 234 | |

### 3.1.2 Jitter

Jitter (measured in microseconds or % jitter) is a common perturbation measure of the cycle-to-cycle frequency variation or instability of a speech signal, expressed as:

$$Jitter = \frac{1}{N-1}\sum_{i=1}^{N-1}|T_i - T_{i+1}| \qquad (2)$$

where $T_i$ is the extracted period of the $i^{th}$ speech signal segment and $N$ is the number of extracted speech signal segments[52].

Studies have found higher jitter measures in smokers compared to non-smokers. Gonzalez and Carpi found jitter increased between non-smokers and smokers of less than 10 years, but the difference was only significant in men (47.67μs non-smoker male vs 62.78 μs smoker male, p<0.05). It is suggesting that changes in jitter may be related to long term smoking[20]. In[53], authors confirmed that the jitter value was higher in women aged 18-24 years who smoked compared to non-smoking women, but the difference was not significant. Women in the smoking group had a relatively short history of smoking (3.5 years on average), which may account for the study findings.

Three studies evaluated the voice changes over different smoking frequencies and smoking histories. In[21], 32 adults (12 smokers without voice problems, 8 smokers with voice problems, and 12 non-smokers) were evaluated by the phonatory tasks. The results of the jitter analyses shown that smokers with voice problems present statistically significant higher jitter values for all speech tasks than non-smokers. In[54], jitter values were significantly higher in smoking males who smoked at least five cigarettes a day for five or more years (0.364% smokers vs 0.283% non-smokers) than in non-smoking males. In an evaluation of voice features comparing women who never smoked with women who smoked less than 10 years, and women who smoked 10 or more years[55] the investigators found that jitter value was increased in women who smoked compared to non-smokers, but only the jitter difference between non-smokers and smokers who smoked 10 or more years was significant (1.11% smoker ≥ 10 years vs 0.92% smoker < 10 years vs 0.69% non-smoker). However, the authors also noted that the fact that women who had a longer smoking habit also smoked more cigarettes per day and were older than the other groups, could account for the difference in voice perturbation.

Table 3 Comparison of smokers' vs non-smokers' jitter

| Author(s) | Methods | Smokers | | Non-smokers | |
|---|---|---|---|---|---|
| *Gonzalez and Carpi (2004)* | Sustained vowels | Male 62.78 | Female 55.11 | Male 47.67 | Female 45.60 |
| *Awan (2011)* | Sustained vowels | Female 0.40 ± 0.17 | | Female 0.37 ± 0.15 | |
| *Guimarães and Abberton (2005)* | % Jitter /a/ /i/ /u/ | Mixed 1.10 0.86 0.73 | | Mixed 0.52 0.47 0.51 | |
| *Chai et al. (2011)* | Sustained Vowels | Male 0.364 | | Male 0.283 | |
| *Vincent and Gilbert (2012)* | Sustained Vowels | < 10 years 0.92 | ≥ 10 years 1.11 | Non-smokers 0.69 | |

### 3.1.3 Shimmer

Shimmer (measured in decibels [dB] or % shimmer) is another common perturbation measure in the acoustic analysis, which is a measure of amplitude variation of a speech signal, can be expressed as:

$$Shimmer = \frac{1}{N-1}\sum_{i=1}^{N-1}|20\log{(A_{i+1}/A_i)}| \qquad (3)$$

where $A_i$ is the extracted peak-to-peak amplitude of the $i^{th}$ cycle of the speech signal and $N$ is the number of extracted cycles of the speech signal[52].

Studies have found higher shimmer values in smokers compared to non-smokers. In one study[54], the percentage of shimmer was significantly higher in male smokers when compared to male non-smokers (4.57% vs 2.50%). Similarly, in another study[55], the shimmer was significantly higher for female smokers who smoked more than 10 years than for either non-smokers and smokers who smoked less than 10 years (0.37 dB smoker ≥ 10 years vs 0.25 dB smoker <10 years vs 0.21 dB non-smoker). Zealouk et al. examined the voice features of 40 male adults, 20 smokers with a median duration of 13 years[56]. Both jitter and shimmer values were significantly higher for smokers when compared to non-smokers (jitter: 51.997 μs smokers vs 36.989 μs non-smokers, p<0.05; shimmer: 0.570 dB smokers vs 0.378 dB non-smokers, p<0.01).

In a study with 81 men, 21 of whom were former cigarette smokers that had been using e-cigarettes for one to three years, 30 were users of conventional cigarettes with a smoking history of one to five years, and 30 were non-smokers[57], the absolute shimmer was significantly different between conventional cigarette smokers and e-cigarette smokers and non-smokers, with increased shimmer in the conventional cigarette users (0.34 dB conventional cigarette vs 0.22 dB e-cigarette smokers vs 0.22 dB non-smokers), however, there was no significant difference between groups for $F_0$ or jitter.

Table 4 Comparison of smokers' vs non-smokers' shimmer

| Author(s) | Methods | Smokers | | Non-smokers |
|---|---|---|---|---|
| *Chai et al. (2011)* | Sustained vowels | Male 4.569 | | Male 2.497 |
| | | < 10 years | >10 years | Female |

| Vincent and Gilbert (2012) | /a/ | 0.31 | 0.38 | 0.23 |
| --- | --- | --- | --- | --- |
| | /i/ | 0.20 | 0.34 | 0.20 |
| | /u/ | 0.20 | 0.36 | 0.18 |
| Zealouk et al. (2018) | | Male | | Male |
| | /a/ | 0.648 | | 0.355 |
| | /i/ | 0.510 | | 0.379 |
| | /u/ | 0.551 | | 0.401 |
| Tuhanioğlu et al. (2019) | Sustained vowels | cigarettes | e-cigarettes | Male |
| | % Shimmer | 3.81±2.71 | 2.60±0.95 | 2.67±0.83 |
| | Shimmer dB | 0.34±0.24 | 0.22±0.08 | 0.22±016 |

### 3.1.4 Harmonics to Noise Ratio (HNR)

Harmonics to noise ratio (HNR), expressed in dB, represents the degree of acoustic periodicity. It is the ratio between a periodic component and a non-periodic component of a speech, which is a measure that quantifies the amount of additive noise in the voice signal. HNR is also used as a measure for the signal-to-noise ratio (SNR) of a periodic signal to determine the voice quality[58]. HNR can be formulated as the following equation according to[59]:

$$HNR = 10 * log_{10} \frac{AC_V(T)}{AC_V(0) - AC_V(T)} \quad (4)$$

where $AC_V(0)$ is the autocorrelation coefficient at the origin consisting in the all energy of the speech signal. The $AC_V(T)$ is the component of the autocorrelation corresponding to the fundamental period.

Although HNR has been labelled as an index of vocal ageing[60], studies have found that HNR was lower among smokers in comparison with non-smokers. In[61], Braun found that the HNR value in the smoker group (9.4dB) was lower than the non-smokers group (11.4dB). Díaz et al. found that the amount of noise presented in the smokers' voices was evidently higher than the amount of noise in the voice of the non-smokers[62].

Studies have also found that the HNR value is affected by the duration of smoking. Pinar et al.[24] evaluated 109 young adult men among whom were 58 smokers (52 of these had smoked for less than ten years). The results indicated the smokers' NHR (25.01dB) was slightly lower than non-smokers' (25.74dB). In a study involving 210 young (average age 22 years) adult females attending smoking status assessment[63], with average years of smoking only 2.16 (SD: ±1.29) and average number of cigarettes smoked daily 13.19 (SD: 6.65), no statistically significant differences were noted for HNR values for smokers with short smoking years but heavy daily smoking habits compared to non-smokers. Pintoa and Crespob investigated HNR as features to classify smokers voice and non-smokers voice[64]. 40 smokers with an average smoking duration of 30 years and 40 non-smokers were measured, the HNR value of smokers was different with that of non-smokers (0.051% vs 0.016%).

Table 5 Comparison of smokers' vs non-smokers' HNR

| Author(s) | Methos | Smokers | Non-smokers |
|---|---|---|---|
| *Braun (2019)* | Oral reading | Male | Male |
| | /a/ | 9.4 | 11.4 |
| *Díaz et al. (2014)* | Sustained | Mixed | Mixed |
| | vowels | 25.22 | 36.1 |
| *Pinar et al. (2016)* | Sustained | Male | Male |
| | vowels | 25.01 | 25.74 |
| *Tafiadis et al.* | | Female | Female |
| *(2017)* | /a/ | 24.65 | 24.94 |
| | /e/ | 25.30 | 25.65 |
| *Pintoa et al. (2011)* | Sustained | Mixed | Mixed |
| | vowels | 0.051 | 0.016 |

### 3.1.5 *Formant frequencies ($F_1$, $F_2$, $F_3$, and $F_4$)*

A formant frequency is a concentration of acoustic energy around a particular frequency in the speech wave, which is a distinctive frequency component of the acoustic signal produced by speech[65]. A formant frequency with the lowest frequency is named $F_1$, the second $F_2$, the third $F_3$, and the fourth $F_4$. Each vowel in English has three formant frequencies and most often, $F_1$ and $F_2$ are enough to determine a vowel.

Zealouk et al. reported that smokers' formant frequencies $F_1$ and $F_2$ were close to those of non-smokers, and smokers' $F_3$ and $F_4$ were lower than that in non-smokers[56]. On the other hand, $F_1$, $F_2$, and $F_3$ values dramatically decreased with age increasing, and these values for men were lower than those for women.

Table 6 Comparison of smokers' vs non-smokers' formant frequencies (Hz)

| Author(s) | Methods | Smokers | | | | Non-smokers | | | |
|---|---|---|---|---|---|---|---|---|---|
| | | $F_1$ | $F_2$ | $F_3$ | $F_4$ | $F_1$ | $F_2$ | $F_3$ | $F_4$ |
| Zealouk et al. (2018) | /a/ | 850 | 1600 | 2600 | 3500 | 900 | 2000 | 3050 | 4100 |
| | /i/ | 400 | 1900 | 2700 | 3900 | 500 | 2100 | 3050 | 4300 |
| | /u/ | 450 | 1500 | 2500 | 3700 | 550 | 1400 | 2950 | 4050 |

### 3.1.6 Other Features

Pitch is the feature to judge sounds as its highness and lowness, which depends on the vibrational frequency produced by the vocal cords during the sound production. Pitch can be quantified using fundamental frequency ($F_0$) as it is correlated with the physical feature of $F_0$[66]. Both pitch and $F_0$ are often used interchangeably in the literature. Nonetheless, a few studies found that smokers had lower pitch values than those of non-smokers[21,56]. In study[56], Zealouk et al. found that the pitch value for smokers was statistically lower than non-smokers.

Correlation dimension (D2) is a nonlinear dynamic quantitative measurement that can be applied to voice signals[67]. Chai et al indicated that D2 values were sensitive to cigarette smoking and smokers had significantly higher D2 value than non-smokers[54].

Table 7 Comparison of smokers' and non-smokers' other features

| Author(s) | Features | Smokers | Non-smokers |
|---|---|---|---|
| Zealouk et al. (2018) | Pitch (Hz) | Male | Male |
| | /a/ | 143 | 168 |
| | /i/ | 140 | 159 |
| | /u/ | 147 | 164 |
| Chai et al. (2011) | | Male | Male |
| | SNR | 18.076 | 21.863 |
| | ERR | 0.324 | 0.03 |
| | D2 | 2.205 | 1.681 |

### 3.2 The Effects of Smoking Cessation on Voice Features

Findings from a large longitudinal study suggest that changes in the fundamental frequency ($F_0$) are reversible when individuals quit smoking. Berg et al. evaluated voice frequency in 2274 adults aged 40-79 years classified as non-smokers, former smokers, and current smokers[68]. Regression analysis found significant differences in $F_0$, but found only marginal differences between former smokers and non-smokers regardless of gender. $F_0$ was significantly lower in current smokers compared to non-smokers (103.7 Hz male smokers vs 112.5 Hz male non-smokers, p<0.001; 159.1 Hz female smokers vs 170.7 Hz female non-smokers p<0.001) and former smokers (103.7 Hz male smokers vs 114.7 Hz male former smokers, p<0.001; 159.1 Hz female smokers vs 166.1 Hz female non-smokers p<0.001). However, the study did not report on the abstinence duration. Therefore, it is unclear how soon changes in $F_0$ can be detected after quitting.

Three studies evaluated changes in voice features over short periods of abstinence. In the first of these studies[22], $F_0$ was measured before, during, and after a 40 hour period of abstinence in two smokers and two non-smokers. There was a small increase in $F_0$ for smokers after 40 hours abstinence, with no changes in $F_0$ for the control subjects. In the second study[69], voice features were measured before abstinence, one-week abstinence, and one-month abstinence in 18 smokers (5 female, 13 male). On average, $F_0$ was higher during abstinence than before abstinence for both males (103.27 Hz pre-abstinence vs 107.08 Hz one-week vs 109.71 Hz one-month) and females (187.37 Hz pre-abstinence vs 192.91 Hz one-week vs 207.19 Hz one-month) but the difference was not significant. In the third study, Dirk and Braun also found a decrease in jitter across the abstinence period, but the difference was only significant between the pre-abstinence and one-month abstinence time points (0.50% pre-abstinence vs 0.21% one month, p=0.00954). There was also a significant decrease in shimmer across the abstinence period, with the largest difference between the pre-abstinence and one-week abstinence time points (5.98% pre-abstinence vs 4.60% one week, p=0.03189; 5.98% vs 4.64% one month, p=0.00988). In another study of 20 female smokers (duration and quantity of smoking not given) before and after smoking cessation for 6 months, the results were compared with 40 age-matched non-smokers[70]. The results found that an increase in $F_0$ was present after Reinke's Edema microsurgery and 6-months of smoking cessation but not fully reversible to normal voice quality (as in non-smokers) due to the vocal alterations caused by smoking.

## 4. Discussion

Fundamental frequency ($F_0$), jitter, and shimmer are the voice features that have been most used to analyse and discriminate between smokers and non-smokers. Harmonics-to-noise ratio (HNR) acts as a supplement for the smoking voice analysis tasks. A few papers have extracted formant frequency, pitch, and correlation dimension (D2) as a measurement. Although there are a number of limitations in the literature (such as small samples, gender imbalance, and limited statistical analysis), there is sufficient evidence to support our contention that smoking consistently affects various voice features in specific ways. We also found evidence that the effects of smoking on voice features such as $F_0$, jitter, and shimmer may be reversed after a period of smoking cessation but are not fully reversible.

Overall, it appears that $F_0$ is affected by smoking in a relatively early stage of smoking history. A period of smoking abstinence would also affect $F_0$, especially in women. Findings also suggest that jitter and shimmer may be sensitive to the duration of smoking history. Significant increases in both jitter and shimmer have been found in studies where subjects have a long history of smoking but are mixed in studies where subjects have a shorter history of smoking. Some studies also suggest that perturbation measures may be sensitive to the number of cigarettes smoked per day.

Table 8 below provides a summary of the strengths and the weaknesses of these main voice features that have been evaluated in relation to smoking. In addition, there is research to suggest a degree of specificity in detecting active versus passive smoking[51], and traditional cigarettes versus e-cigarettes, non-combustible tobacco, and water pip smoking[57,71,72].

Table 8. A summary of the voice features

| Features | Advantages | Disadvantages |
| --- | --- | --- |
| Fundamental frequency ($F_0$) | Sensitivity to smoking, especially for heavy smokers<br>Specificity for active smoking<br>Reversal with long-term abstinence | Affected by age and other factors<br>Not a valid criterion to distinguish all smokers |
| Jitter | Sensitivity to smoking cessation<br>Specificity for combustible tobacco | Changes slower over a longer period of time<br>Affected by voice disorders |

|  |  | Sensitivity to the length of smoking history |
|---|---|---|
| Shimmer | Significantly affected by smoking | Sensitivity to the duration of smoking |
|  | Sensitivity to smoking cessation |  |
| HNR | Significantly affected by smoking | Changes slower over a longer period |
|  | Sensitivity to smoking cessation | Changes with ageing |
| Formant frequencies | Sensitivity to gender differences | Affected by age |
|  |  | Degrades during signal transmission |
|  |  | Affected by human noise |

More work is needed to develop automated speech assessment for smoking status based on voice features as an alternative objective smoking status validation method to the point where a computer provided with the voice features extracted from speech recordings of smokers and non-smokers may be able to discriminate whether a given speech sample is from a smoker or non-smoker. Automated speech assessment for smoking status would have many advantages, including quantitative and objective assessment, able to be performed remotely, reducing analysis cost and time, and readily integrated into screening and remote health monitoring applications. In a future project, we will use $F_0$, jitter, and shimmer to distinguish smokers from non-smokers by applying combinations of speech signal processing and machine learning techniques.

Although this literature review has revealed the effects of smoking on different voice features of speech, the relationship of these voice features with speaker smoking behaviour is complex. Smoking frequency, smoking history (duration and type of product, including the use of newer products such as e-cigarettes and heat-not-burn tobacco devices) need to be considered. Other factors, such as age, gender, presence of chronic respiratory disease, use of inhaled steroids, and alcohol use should also be accounted for. We found limited data on the length of time after quitting smoking that it takes to measure a change, and the level of smoking reduction required to create a change. An important issue is the lack of a reference database to establish the methodologies for smoking status classification from speech signal analysis. We aim to build a long-term smoking cessation voice recording corpus based on this study and implement our voice-based smoking status validation model in a mobile application to provide an objective self-report measure method in smoking cessation trials and clinical practice.

## 5. Conclusion

In this study, we conducted a comprehensive investigation of the effects of voice features in the detection of smoker/non-smoker speech signal. This paper has presented a comparative review of smoker's voice features affected by smoking. We conclude that acoustic voice parameters appear to be influenced by smoking and smoking cessation: Fundamental frequency ($F_0$), jitter, shimmer, and harmonics to noise ratio (HNR) are affected by cigarette smoking. Smokers have a lower fundamental frequency than non-smokers in both gender and age groups. Smokers present higher jitter values for all vowels. Smokers' shimmer values are higher than the values of non-smokers. During smoking cessation, HNR value increase dramatically. Moreover, jitter and shimmer decrease significantly. $F_0$ value rises during smoking abstinence and decreases again after resuming smoking. However, more research with larger samples is needed to refine the sensitivity and specificity of this approach to be able to translate it into a real-time tool.